\documentclass[aps,prl,twocolumn,superscriptaddress]{revtex4}%
\usepackage{amssymb}
\usepackage{epsfig}
\usepackage{graphicx,amsmath,bm}%
\usepackage{color}
\usepackage{amsmath}%
\usepackage{amsfonts}%
\usepackage{graphicx}

\begin{document}
\title{Dissipation-induced d-Wave Pairing of Fermionic Atoms in an Optical Lattice}
\author{S. Diehl}
\affiliation{Institute for Quantum Optics and Quantum Information of the Austrian Academy
of Sciences, A-6020 Innsbruck, Austria}
\affiliation{Institute for Theoretical Physics, University of Innsbruck, A-6020 Innsbruck, Austria}
\author{W. Yi}
\email{wyiz@ustc.edu.cn}
\affiliation{Institute for Quantum Optics and Quantum Information of the Austrian Academy
of Sciences, A-6020 Innsbruck, Austria}
\affiliation{Key Laboratory of Quantum Information, University of Science and Technology of China,
CAS, Hefei, Anhui,230026, People's Republic of China}
\author{A. J. Daley}
\affiliation{Institute for Quantum Optics and Quantum Information of the Austrian Academy
of Sciences, A-6020 Innsbruck, Austria}
\affiliation{Institute for Theoretical Physics, University of Innsbruck, A-6020 Innsbruck, Austria}
\author{P. Zoller}
\affiliation{Institute for Quantum Optics and Quantum Information of the Austrian Academy
of Sciences, A-6020 Innsbruck, Austria}
\affiliation{Institute for Theoretical Physics, University of Innsbruck, A-6020 Innsbruck, Austria}

\begin{abstract}
We show how dissipative dynamics can give rise to pairing for two-component fermions on a lattice.  In particular, we construct a ``parent''  Liouvillian operator so that  a BCS-type state of a given symmetry, e.g. a d-wave state, is reached for arbitrary initial states in the absence of conservative forces.  The system-bath couplings describe single-particle, number conserving and quasi-local processes. The pairing mechanism crucially relies on Fermi statistics. We show how such Liouvillians can be realized via reservoir engineering with cold atoms representing a driven dissipative dynamics.
\end{abstract}

\maketitle
\affiliation{\mbox{\it ${}^a$Institute for Quantum Optics and Quantum
Information of the Austrian Academy of Sciences,}}
\affiliation{\mbox{\it A-6020 Innsbruck, Austria}}

Pairing in condensed matter physics in general, and in atomic quantum gases in
particular, is associated with conservative forces between particles, e.g., in
Cooper pairs or molecular BEC pairs \cite{leggettbook}. Lattice dynamics gives rise to exotic forms of pairing, such as the expected formation of d-wave Cooper pairs of
fermions for a 2D Hubbard model for repulsive interactions, as discussed in
the context of high-$T_{c}$ superconductivity \cite{Anderson87}, but also condensates of $\eta$-pairs \cite{YangEta}, and the formation of repulsively bound atom pairs \cite{RepPairs}. Here we show
that purely dissipative dynamics, induced by coupling the system to a bath,
can give rise to pairing, even in the complete absence of conservative forces. This \textquotedblleft dissipative pairing\textquotedblright crucially relies on Fermi statistics and is in contrast to
pairing arising from bath-mediated interactions (e.g., phonon-mediated Cooper
pairing). We will discuss how reservoir engineering provides
opportunities for experimental realisation of this dissipative pairing mechanism with cold atomic fermions in optical lattices \cite{Ketterle06}.

Below we treat the example of a d-wave-paired BCS state of two-component
fermions in two dimensions (2D), showing how the pairing can be generated via purely
\textit{dissipative} processes. A BCS-type state is the conceptually simplest
many-body wave function describing a condensate of $N$ paired spin-1/2
fermionic particles, $|\mathrm{BCS}_{N}\rangle\sim(d^{\dag})^{N/2}%
|$vac$\rangle$. On a square lattice, and assuming singlet pairs with zero
center-of-mass momentum, we have $d^{\dag}\hspace{-0.07cm}=\sum_{\mathbf{q}%
}\varphi_{\mathbf{q}}c_{\mathbf{q},\uparrow}^{\dag}c_{-\mathbf{q},\downarrow
}^{\dag}$ or $d^{\dag}=\sum_{i,j}\varphi_{ij}c_{i,\uparrow}^{\dagger
}c_{j,\downarrow}^{\dagger}$, where $c_{\mathbf{q},\sigma}^{\dagger}$ ($c_{i,\sigma
}^{\dagger}$) \ denotes the creation operator for fermions with quasimomentum
$\mathbf{q}$ (on lattice site $i$) and spin $\sigma=\uparrow,\downarrow$, and
$\varphi_{\mathbf{q}}$ ($\varphi_{ij}$) the momentum (position)\ wave function
of the pairs. For d-wave pairing, the pair wave function obeys
$\varphi_{q_{x},q_{y}}=-\varphi_{-q_{y},q_{x}}=\varphi_{-q_{x},-q_{y}}$, and below we choose
$\varphi_{\mathbf{q}}=\cos q_{x}-\cos q_{y}$ or $\varphi_{ij}= \tfrac{1}{2}\sum
_{\lambda=x,y}\rho_\lambda(\delta_{i,j+\mathbf{e}_{\lambda}}+  \delta_{i,j-\mathbf{e}%
_{\lambda}})$ with $\rho_x =-\rho_y=1$ corresponding to the limit of well localized pairs (see Fig.~1a), and $\mathbf{e}_{\lambda}$ the unit lattice vector in $\lambda =x,y$ direction.
For reference below we remark that in BCS theory, with pairing induced by
\textit{coherent} interactions, the corresponding energy gap function would be
$\Delta_{\mathbf{q}}=\Delta\left(  \cos q_{x}-\cos q_{y}\right)  $ in the
molecular limit. The dissipative pairing mechanism is readily generalized to other pairing symmetries, such as e.g. $p_x + \mathrm i p_y$ \cite{Gurarie05}, as long as the pairing is not onsite.

\begin{figure}[tb]
\includegraphics[width=8cm]{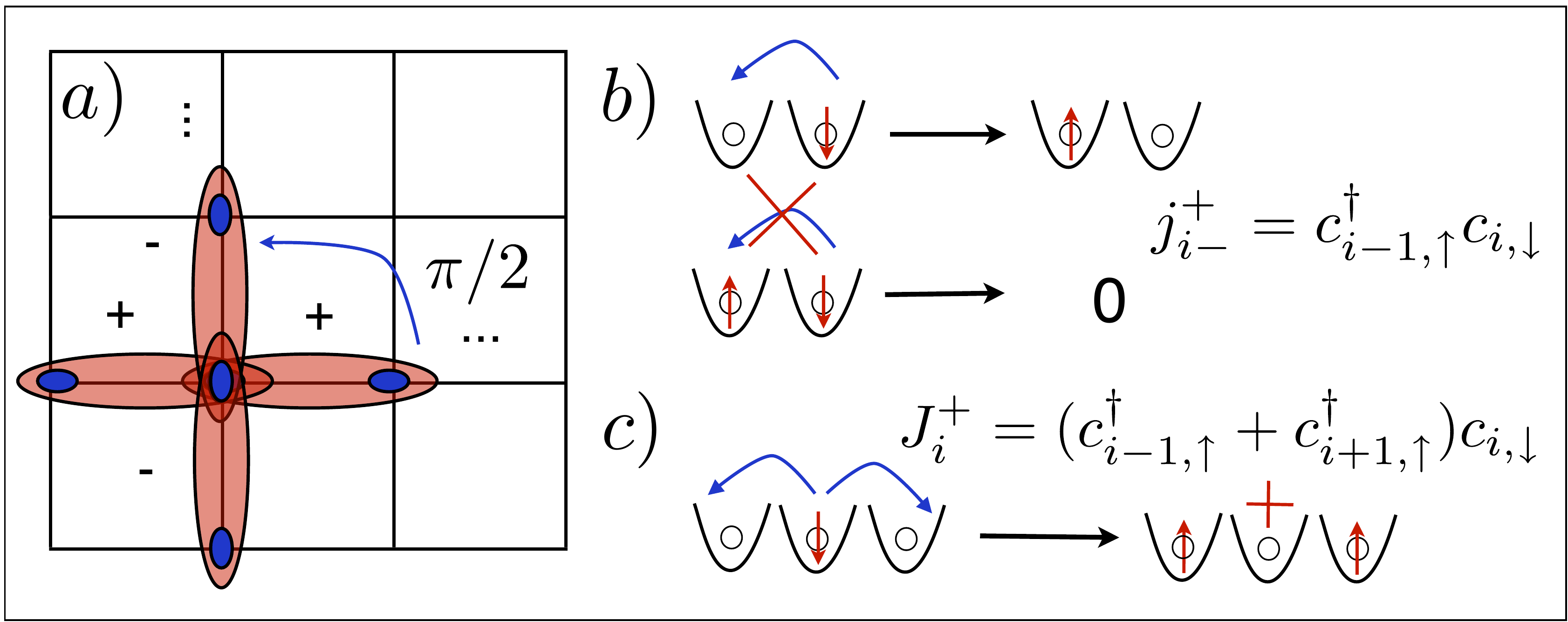}
\caption{(a) Symmetry in the
d-wave state, represented by a single offsite fermion pair exhibiting the
characteristic sign change under spatial rotations. In a d-wave BCS state,
this pair is delocalized over the whole lattice. (b,c) The dissipative pairing
mechanism builds on (b) Pauli blocking and (c) delocalization via phase
locking. (b) Illustration of the action of Lindblad operators using Pauli
blocking for a N\'eel state (see text). (c) The d-wave state may be seen as a
delocalization of these pairs away from half filling (shown is a cut along one lattice axis). }%
\label{Fig1_Jump_Operators}%
\end{figure}

While in the standard scenario BCS-type states are typically used as
variational mean-field wavefunctions to describe pairing due to interactions,
here the system is dissipatively driven towards the (pure) many-body BCS
state, $\rho(t)=e^{\mathcal{L}t}\rho(0)\overset{t\rightarrow\infty
}{\longrightarrow}|\mathrm{BCS}_{N}\rangle\langle\mathrm{BCS}_{N}|$, beginning
from an arbitrary initial mixed state $\rho(0)$. The dynamics of the density
matrix for the $N$-particle system $\rho(t)$ is generated by a Liouville
operator with the structure $\mathcal{L}\rho=-\mathrm{i} H_{\mathrm{eff}}\rho+\mathrm{i}\rho
H_{\mathrm{eff}}^{\dagger}+\kappa\sum_{\ell}j_{\ell}\rho j_{\ell}^{\dagger}$
with non-hermitian effective Hamiltonian $H_{\mathrm{eff}}=H-\frac{\mathrm i}%
{2}\kappa\sum_{\ell}j_{\ell}^{\dagger}j_{\ell}$. Here, $\{j_{\ell}\}$ are
non-hermitian Lindblad operators reflecting the system-bath coupling with
strength characterized by the rate $\kappa$. The Hamiltonian $H$ generates
unitary evolution, and will be set to zero for most of the discussion. The pure paired BCS state
being the unique steady state of the dissipative dynamics results from the
possibility to identify a set of operators with $j_{\ell}|\mathrm{BCS}%
_{N}\rangle=0$ $\forall\ell$ \cite{Diehl08,Verstraete09}.

Below, we will identify these operators $j_{\ell}$ for the d-wave paired BCS
states, and in addition study the dynamics close to the final steady state,
i.e., near $|\mathrm{BCS}_{N}\rangle$. We can then investigate the complex
excitation spectrum of $\mathcal{L}$, where, remarkably, we find a dissipative
\textquotedblleft BCS gap\textquotedblright\ that implies exponential approach
to the steady state.

We can readily check that the Lindblad operators $j_{\ell}$ generating the
d-wave BCS state are given by
\begin{eqnarray}\label{DWjump}
J_{i}^{\alpha}=\sum_{\lambda=x,y}\rho_\lambda(c_{i+\mathbf{e}_{\lambda}%
}^{\dagger}+c_{i-\mathbf{e}_{\lambda}}^{\dagger})\sigma^{\alpha}c_{i},
\end{eqnarray}
with 2-spinor $c_{i}=(c_{i,\uparrow},c_{i,\downarrow})^{T}$ and $\sigma^{\alpha}$ Pauli matrices with $\alpha=\pm,z$ or $\alpha=x,y,z$. An explicit
construction is given below. Remarkably, these Lindblad
operators, which generate pairing dissipatively, are bilinear and number conserving, thus acting on a  \textit{single-particle} only. They are
also quasi-local operators, involving only a plaquette of nearest neighbor
sites (see Fig.~1a).

Before entering the more technical discussion of obtaining these
$J_{i}^{\alpha}$, we discuss the dynamics for states close to the final state
$|\mathrm{BCS}_{N}\rangle$, where the physics is particularly transparent and
analogies to the usual case of interaction-induced pairing in BCS theory can
be made. For states close to $|\mathrm{BCS}_{N}\rangle$ we can linearize the
master equation dynamics using a Bogoliubov-type approach. Here we take advantage of the fact that we know the steady state for our problem exactly; this knowledge can be used to construct a quadratic theory for the fluctuations on top of it. For this purpose it
is technically convenient to give up exact particle number conservation, and to work with
fixed phase coherent states $|\mathrm{BCS}_{\theta}\rangle=\mathcal{N}%
^{-1/2}\exp(e^{\mathrm{i}\theta}d^{\dag})|\text{vac}\rangle$ instead of the
number states $|\mathrm{BCS}_{N}\rangle$ \cite{leggettbook}, where $\mathcal{N}%
=\prod_{\mathbf{q}}(1+\varphi_{\mathbf{q}}^{2})$ ensures the normalization.
The density matrix for these states, describing the dark steady state, factorises in momentum space since $\exp(e^{\mathrm{i}%
\theta}d^{\dag})|\text{vac}\rangle=\prod_{\mathbf{q}}(1+e^{\mathrm{i}%
\theta}\varphi_{\mathbf{q}}c_{\mathbf{q},\uparrow}^{\dag}c_{-\mathbf{q}%
,\downarrow}^{\dag})|\text{vac}\rangle$. At late times, we can therefore expand the state around $|\mathrm{BCS}_{\theta}\rangle$ by making the factorized ansatz $\rho=\prod_{\mathbf{q}}\rho_{\mathbf{q}}$, where $\rho_{\mathbf{q}}$ contains the modes $\pm(\mathbf{q},\sigma)$ necessary
to describe pairing. We can then utilize the projection prescription
$\rho_{\mathbf{q}}=\mathrm{tr}_{\neq\mathbf{q}}\rho$ to find the
equations of motion for the single pair density matrices $\rho_{\mathbf{q}}$
in the presence of nonzero mean fields. These result from the coupling to other
momentum modes, and their values are dictated by the final state properties.
The resulting effective Hamiltonian is quadratic:
\begin{eqnarray}
H_{\mathrm{eff}}  & =&-\tfrac{\mathrm{i}\kappa}{2}\sum_{\mathbf{q},\sigma
}\big\{\tilde n (c_{\mathbf{q},\sigma}^{\dag}c_{\mathbf{q},\sigma} + |\varphi_{\mathbf{q}}|^{2} c_{\mathbf{q},\sigma}c_{\mathbf{q},\sigma}^{\dag})\\\nonumber
&&\quad +\tilde \Delta_{\mathbf{q}}s_\sigma c_{-(\mathbf{q},\sigma)} c_{\mathbf{q},\sigma}+\text{h.c.}\big\}
 =-\tfrac{\mathrm{i}}{2}\sum_{\mathbf{q},\sigma}\kappa_{\mathbf{q}}%
\gamma_{\mathbf{q},\sigma}^{\dag}\gamma_{\mathbf{q},\sigma},%
\end{eqnarray}
with $s_\uparrow = -1,s_\downarrow =1$ and dimensionless ``gap function" $\tilde \Delta_\mathbf{q} = \tilde \Delta\varphi_\mathbf{q}$, and where the diagonal and off diagonal mean fields evaluate to $\tilde n=  |\tilde \Delta| =2\int\tfrac{d\mathbf{q}}{(2\pi)^2}\tfrac{|\varphi_\mathbf{q}|^2}{1 +|\varphi_\mathbf{q}|^2} \approx 0.72$ on the d-wave state, where the integration is over the Brillouin zone. We diagonalize $H_{\mathrm{eff}}$ in the second line, introducing quasiparticle Lindblad operators%
\[
\gamma_{\mathbf{q},\sigma}=(1+\varphi_{\mathbf{q}}^{2})^{-1/2}\,(c_{-\mathbf{q},\sigma}+s_\sigma\varphi_{\mathbf{q}}c_{\mathbf{q},-\sigma
}^{\dag}).
\]
In this basis, the resulting master equation reads $\partial_t\rho = -iH_{\mathrm{eff}}\rho+i\rho
H_{\mathrm{eff}}^{\dagger}+\sum_{\mathbf{q},\sigma} \kappa_{\mathbf{q}} \gamma_{\mathbf{q},\sigma}\rho \gamma^\dag_{\mathbf{q},\sigma}$. The linearized Lindblad operators have analogous properties to quasiparticle operators familiar
from interaction pairing problems: (i) They annihilate the (unique) steady
state $\gamma_{\mathbf{q},\sigma}|\mathrm{BCS}_{\theta}\rangle=0$; (ii) they
obey the Dirac algebra $\{\gamma_{\mathbf{q},\sigma},\gamma_{\mathbf{q}%
^{\prime},\sigma^{\prime}}^{\dag}\}=\delta_{\mathbf{q},\mathbf{q}^{\prime}%
}\delta_{\sigma,\sigma^{\prime}}$ and zero otherwise \footnote{The simple
algebra emerging at late times contrasts with the properties of $J_{i}%
^{\alpha}$, which do not exhibit such a property.}; and (iii) therefore are related
to the original fermions via a canonical transformation. The imaginary
spectrum of the effective Hamiltonian features a ``dissipative pairing gap"
\[
\kappa_{\mathbf{q}}=\kappa\, \tilde n \,(1+\varphi_{\mathbf{q}}^{2})\geq
\kappa\,\tilde n.
\]
The dissipative gap implies an exponential approach to the steady d-wave BCS
state for long times. This can be most easily seen in a quantum trajectory
representation of the master equation, where the system's time evolution
is described by a stochastic wavefunction $|\psi(t)\rangle$ evolving under a non-hermitian Hamiltonian $|\psi(t)\rangle
=e^{-iH_{\mathrm{eff}}t}|\psi(0)\rangle/\left\Vert \ldots\right\Vert $
interrupted with rate $\kappa\left\Vert j_{\ell}|\psi(t)\rangle\right\Vert
^{2}$ by quantum jumps $|\psi(t)\rangle\rightarrow j_{\ell}|\psi
(t)\rangle/\left\Vert \ldots\right\Vert $ so that $\rho(t)=\langle
|\psi(t)\rangle\langle\psi(t)|\rangle_{\mathrm{stoch}}$ (see, e.g., \cite{zollerbook}). We thus see that (i)
the BCS state is a ``dark state" of the dissipative dynamics in the sense
that $j_{\ell}|\mathrm{BCS}_{N}\rangle=0$ implies that there will never be
quantum jump, i.e. the state remains in $|\mathrm{BCS}_{N}\rangle$, and (ii)
states near $|\mathrm{BCS}_{N}\rangle$ show an exponential decay according to
the dissipative gap. Note that it is in marked contrast to dissipative preparation of a non-interacting BEC state in bosonic systems, where an approach polynomial in time is expected \cite{Diehl08}.

\begin{figure}[ptb]
\includegraphics[width=8cm]{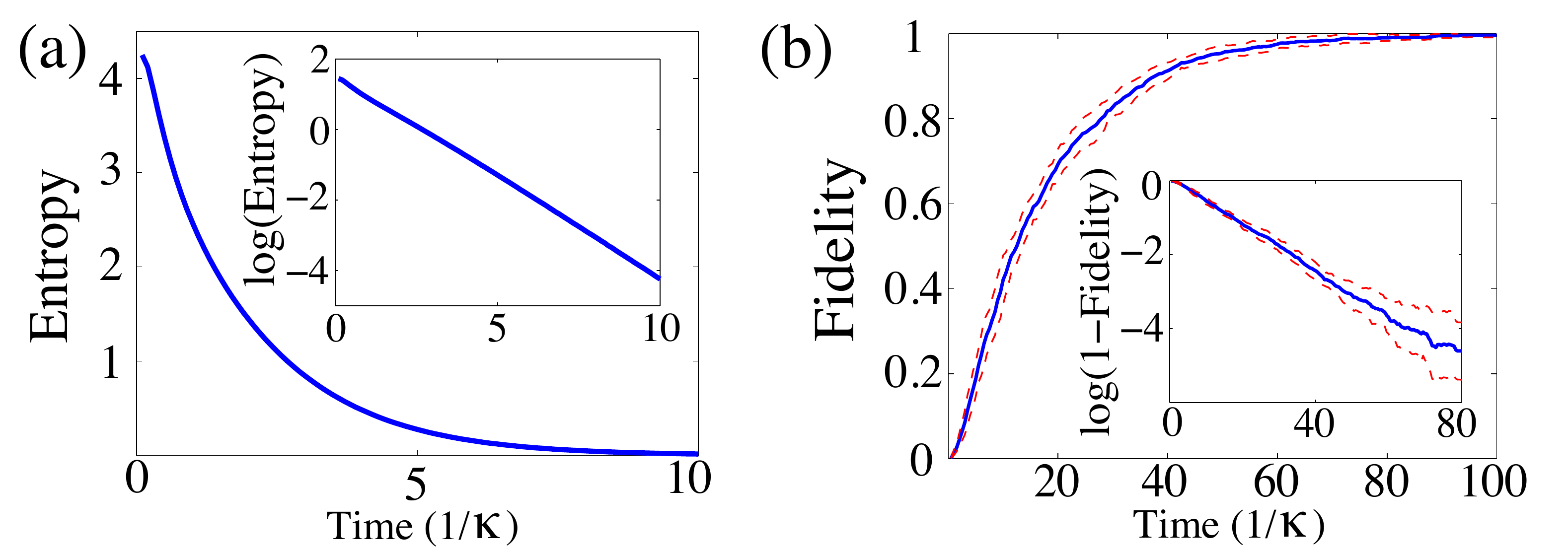}
\caption[Fig. 2]{Numerical illustration of the uniqueness of the steady state, showing evolution under the master equation with Lindblad operators from Eq.~\eqref{DWjump}. (a) Entropy computed exactly for four atoms on a 4x1 lattice, showing exponential convergence from a completely mixed state to a pure state. (b) Fidelity to the d-wave BCS state, $\langle {\rm BCS}_N |\rho |{\rm BCS}_N\rangle$ with 4 atoms on a 4$\times$4 grid, computed via a quantum trajectories method (see text). Dashed lines show sampling error.}%
\label{AFEvol}%
\end{figure}

This convergence to a unique pure state is illustrated in Fig. \ref{AFEvol} using numerical simulations for small systems. In Fig.~2a we show the entropy of the full density matrix for a small 1D system as a function of time, and in Fig.~2b the fidelity of the BCS state for a small 2D grid, computed via the quantum trajectories method.


\emph{Lindblad operators for d-wave states} -- We now turn to the construction
of the Lindblad operators for the d-wave BCS state as given in Eq. \eqref{DWjump}. We
will perform this construction first for an antiferromagnetic N\'{e}el state
at half filling, and then generalize to the BCS state. \ Our task can be
formulated as finding for a given many-body state $|$d$\rangle$ a set of
(non-hermitian) Lindblad operators $j_{\ell}$ so that it becomes the unique
dark state,  $j_{\ell}|\text{d}\rangle=0$ $\forall l$. Both the N\'{e}el and
the BCS state have  product form, $|\text{d}\rangle=\prod_{m}d_{m
}^{\dag}|\text{vac}\rangle$. Thus, we note as a sufficient dark state
condition $[j_{\ell},d_{m}^{\dag}]=0$.

There are two antiferromagnetic N\'{e}el states at half filling $|\text{N}%
+\rangle=\prod_{i\in A}c_{i+\mathbf{e}_{x},\uparrow}^{\dag}c_{i,\downarrow
}^{\dag}|\text{vac}\rangle$, $|\text{N}-\rangle=\prod_{i\in A}c_{i+\mathbf{e}_{x},\downarrow
}^{\dag}c_{i,\uparrow}^{\dag}|\text{vac}\rangle$ with $A$ a
sublattice in a two-dimensional bipartite (square) lattice, which differ by an
overall spin flip. Introducing \textquotedblleft N\'{e}el unit cell
operators\textquotedblright\ $\hat{S}_{i,\nu}^{a}%
=c_{i+\mathbf{e}_{\nu}}^{\dag}\sigma^{a}c_{i}^{\dag}$, $a=\pm,\mathbf{e}_\nu=\{\pm\mathbf{e}_x,\pm\mathbf{e}_y\}$, whose usefulness will become apparent soon, the state can be written in eight
different forms, $|\text{N}\pm\rangle=\prod_{i\in A}\hat{S}_{i,\nu}^{\pm}|\text{vac}\rangle=(-1)^{M/2}\prod_{i\in B}\hat{S}_{i,-\nu}^{\mp}|\text{vac}\rangle$, with $M$ the lattice size. We then see that the Lindblad operators must obey $[j_{i,\nu}^{a},\hat{S}_{j,\mu}%
^{b}]=0$ for all $i,j$ located on the same sublattice $A$ or sublattice $B$, which is
fulfilled for the set
\begin{eqnarray}\label{AFjump}
j_{i,\nu}^{a}=c_{i+\mathbf{e}_{\nu}}^{\dag} \sigma^{a}c_{i}, \, i \in A\,\text{or}\, B.
\end{eqnarray}
Note that these operators can be obtained from $\hat{S}_{i,\nu}^{a}$ by a particle-hole transformation $c_{i,\sigma}^{\dag
}\rightarrow c_{i,\sigma}$ on the central site $i$. For the action of the
operators $j_{i,\nu}^{a}$ the assumption of fermionic
statistics is essential, as illustrated in Fig.~1b: they generate spin
flipping transport according to e.g. $j_{i,\nu}%
^{+}=c_{i+\mathbf{e}_{\nu},\uparrow}^{\dag}c_{i,\downarrow}$, which is not
possible when the antiferromagnetic order is already present. The proof of
uniqueness of the N\'{e}el steady state up to double degeneracy is then
trivial: The steady state must fulfill the quasilocal condition that for any
site occupied by a certain spin, its neighboring sites be filled by
opposite spins. For half filling, the only states with this property are
$|\text{N}\pm\rangle$. The residual degeneracy can be lifted by
adding a single operator $j_{i}=c_{i+\mathbf{e}_{\nu}}^{\dag}%
(\mathbf{1}+\sigma^{z})c_{i}$ at arbitrary $i$.

To find the Lindblad operators for the d-wave BCS state, we apply a similar
strategy. We first rewrite the d-wave generator using the operators $\hat
{S}_{i}^{a}$,
\begin{align}
d^{\dag}   =  \tfrac{\mathrm i}{2}\sum_{i}(c_{i+\mathbf{e}_{x}}^{\dag}&-c_{i+\mathbf{e}_{y}}^{\dag
})\sigma^{y}c_{i}^{\dag}= \tfrac{a}{2}\sum_{i}\hat{D}_{i}^{a},\nonumber\\
\hat{D}_{i}^{a}  & =\sum_{\nu}\rho_\nu\hat{S}_{i,\nu}^{a},\nonumber
\end{align}
where $\rho_{\pm x}=-\rho_{\pm y}=1$, and the quasilocal d-wave pair $\hat D_{i}^{a}$ may be seen as the ``d-wave unit cell operators". Note the freedom of choosing $a=\pm$ in writing the state.
This form makes the physical picture of a d-wave superfluid as delocalized
antiferromagnetic order away from half filling
\cite{Anderson87} particularly apparent. The condition
$[J_{i}^{\alpha},\sum_{j}\hat{D}_{j}^{b}]=0$ ($\alpha=(a,z))$ is fulfilled by
\[
J_{i}^{a}=\sum_{\nu}\rho_\nu j_{i,\nu}%
^{a},\,\,J_{i}^{z}=\sum_{\nu}\rho_\nu j_{i,\nu}^{z},
\]
with $j_{i,\nu}^{z}=c_{i+\mathbf{e}_{\nu}}^{\dag }\sigma^{z}c_{i}$, establishing Eq. \eqref{DWjump}. Similar to above, each
$J_{i}^{a}$ is obtained from $\hat{D}_{i}^{a}$ by a particle-hole
transformation on the central site $i$. In fact, for these operators the
stronger quasi-local commutation properties with the molecular d-wave pairs
holds due to Eq. \eqref{AFjump}: $[J_{i}^{a},\hat{D}_{j}^{a}]=0$ for all
$i,j$, $[J_{i}^{a},\hat{D}_{j}^{b}]=0$ for all
$i,j$ in the same sublattice, which relies again on fermionic statistics. In
contrast, the operators $J_{i}^{z}$ only commute with the symmetric
superposition of all d-wave pairs $\hat{D}_{j}^{a}$. These operators establish
coherence via phase locking between adjacent cloverleaves of sites.

The dark state uniqueness for the Lindblad operators \eqref{DWjump} is
equivalent to the uniqueness of the ground state of the associated hermitian
Hamiltonian $H=V\sum_{i,\alpha=\pm,z}J_{i}^{\alpha \dag}J_{i}^{\alpha}$ for $V>0$. We note
that our BCS state shares  the symmetries of $H$ of global phase
and spin rotations, and translation invariance. Based on the reasonable assumption that no other
symmetries exist, we then expect the ground state to be unique. Note, however, the necessity of the full set $\{J_i^\alpha\}$: Omitting e.g. $\{J_i^z\}$ gives rise to an additional discrete symmetry in $H$ resulting in ground state degeneracy. These results are confirmed with
numerical simulations for small systems and periodic boundary
conditions, as shown in Fig. \ref{AFEvol} .

The above construction method allow us to find ``parent'' Lindblad
operators for a much wider class of BCS-type states. For example, for a $p_x+\mathrm{i}p_y$-wave state of spinless fermions, generated by $p^\dag \sim \sum_{i,\nu}\rho_\nu c^\dag_{i+\mathbf{e}_\nu}c^\dag_{i}$ with $\rho_{x}=-\rho_{-x}=-\mathrm i\rho_{y}$ $= \mathrm i \rho_{-y} =1$, the Lindblad operators are $J_i =\sum_\nu \rho_\nu c^\dag_{i+\mathbf{e}_\nu}c_{i}$.  More generally, they can be obtained for any fixed number pairing state with bilocal pairing \footnote{In one dimension, these states can be parameterized as $|\mu,n,k; N\rangle= O_{k,n,\mu}^{\dag\, N}|\text{vac}\rangle$, where $O^{\dag}_{k,n,\mu} = \sum_{i} \exp\mathrm{i }k x_{i} \,\, c^{\dag}_{i + n}\tau^{\mu} c^{\dag}_{i}$ and $\tau^{\mu} = (\mathbf{1},\sigma^{\alpha})$ with quantum numbers $\mu= 0, ... , 3$, "pairing distance" $n= (1, ... , M-1)$, and pairing momentum $k = ( -(M-1)/2 , ... , (M-1)/2) 2\pi/M$. See W. Yi \emph{et al.}, to be published.}. Note, however, that the construction is not applicable for the onsite (singlet) pairing states -- the analogs of Eq. \eqref{DWjump} become local, such that the lattice sites decouple and no phase coherence can be built up.

\emph{Physical Implementation} -- The quasilocal and number-conserving form of $J_{i}%
^{\alpha}$ raises the possibility to realise dissipative pairing via reservoir
engineering with cold atoms. We illustrate this, considering alkaline earth-like atoms \cite{AEatom1} with nuclear spin (e.g., $I=1/2$ for $^{171}$Yb), and a metastable $^{3}$P$_{0}$ manifold which can be trapped independently to the ground $^{1}$S$_{0}$ manifold. 
In this setting, one can construct a stroboscopic implementation, where the action of each $J_{i}^{\alpha}$ is realised successively. For clarity, we
present this initially in 1D, and choose the example of $J_{i}^+ = (c_{i+
1,\uparrow}^{\dag}+ c_{i- 1,\uparrow}^{\dag})c_{\downarrow}$. The implementation
is depicted in Fig.~3: (i) The $^{3}$P$_{0}$ state
is trapped in a lattice of three times the period as that for the
$^{1}$S$_{0}$ state, defining blocks of three sites in the $^1$S$_0$ lattice.
Using this, any $\downarrow$ atom in $^{1}$S$_{0}$ on the central site
is excited to the $\uparrow$ state of the $^{3}$P$_{0}$ manifold. (ii) By adding an additional potential the traps for $^{3}$P$_{0}$ are divided so that atoms confined in them overlap the right and left sites of
the 3-site block for $^1$S$_0$. (iii) Dissipation is induced via spontaneous decay, obtained by coupling atoms in
the $^{3}$P$_{0}$ state off-resonantly to the $^{1}$P$_{1}$ state, as depicted
in Fig.~3a, with coupling strength $\Omega$, and detuning $\Delta$. If we
couple the $^{1}$S$_{0}$--$^{1}$P$_{1}$ transition to a cavity mode with
linewidth $\Gamma$ and vacuum Rabi frequency $g$, then the decay will be
coherent over the triple of sites. In the limit $\Delta\gg\Omega$ and
$\Gamma\gg\frac{\Omega g}{\Delta}$, we obtain an effective decay rate
$\Gamma_{\mathrm{eff}}=\frac{\Omega^{2}g^{2}}{\Delta^{2}\Gamma}\sim9$kHz for
typical parameters, which bounds the effective dissipative rate for the stroboscopic process, $\kappa$. Provided atoms remain in the lowest band, Fermi statistics will be respected, and coherent dynamics during this process can be neglected in a deep lattice for small scattering lengths.

This operation can occur in parallel for different 3-site blocks, and should be repeated with the superlattice shifted for other
central sites. Similar operations combined with rotations of the nuclear spin before and after these operations allows implementation
of $J_{i}^{-}$ and $J_{i}^{z}$. In 2D 3x3 plaquettes are
defined by the appropriate superlattice potential for the $^{3}$P$_{0}$ level, and the adiabatic manipulation of the potential in
step (ii), should be adjusted to ensure that the correct relative phases are obtained for atoms transported in
orthogonal directions.

\begin{figure}[ptb]
\includegraphics[width=8.5cm]{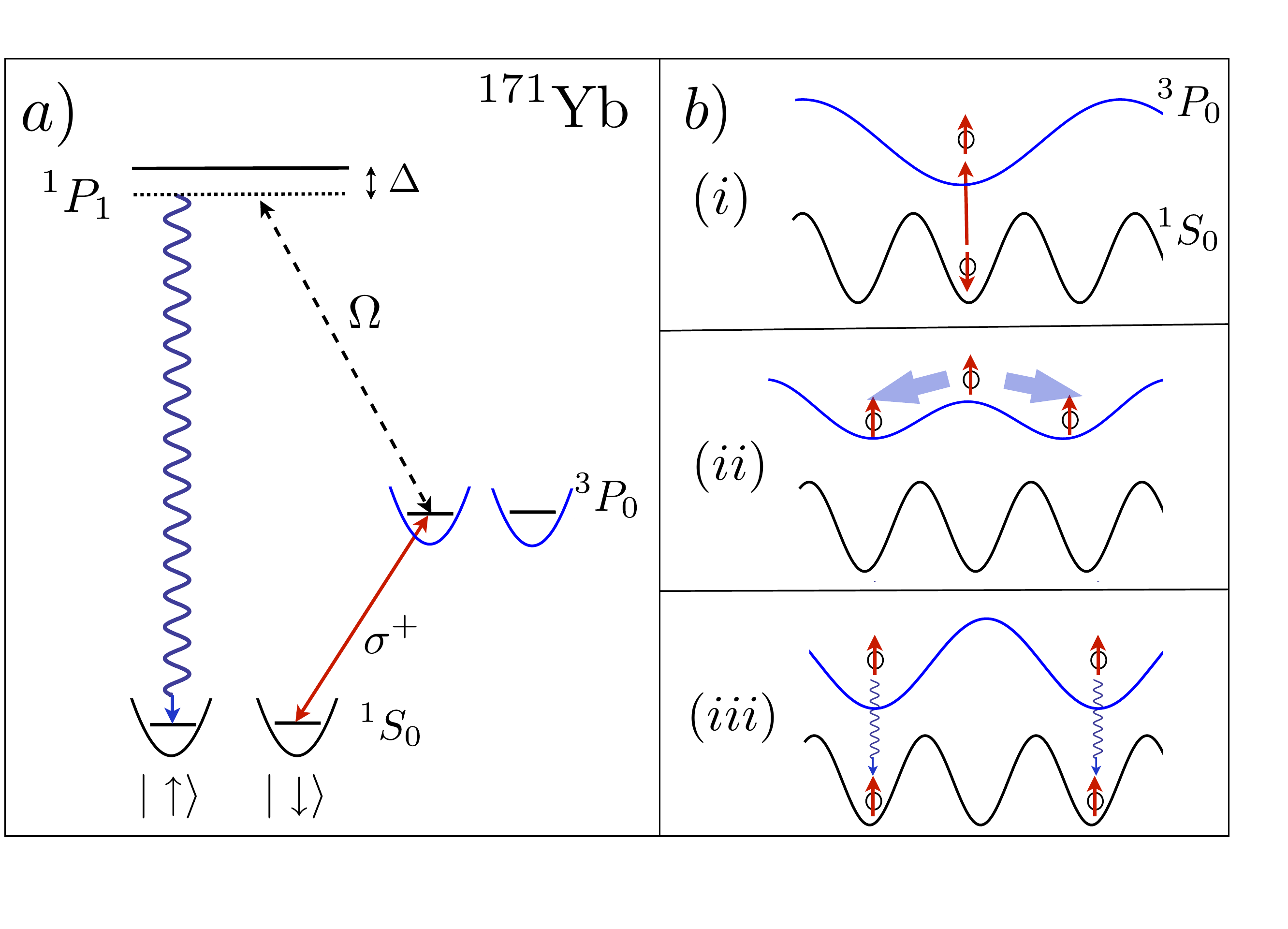}\caption{ (a) Level scheme
for alkaline earth atoms with $I=1/2$, showing excitation to a metastable
level whilst flipping the nuclear spin, and induced decay by coupling to the
$^{1}$P$_{1}$ level.
b) Illustration of $J_{i}^+$ implementation in 1D: (i) A longer period lattice for $^{3}$P$_{0}$ identifies a triple of wells, and atoms from the central level are transferred to the
$^{3}$P$_{0}$ manifold with spin flip. (ii) The $^3$P$_0$ potential wells are adiabatically split
 into two; (iii) Decay is induced,
returning the atom to the $^{1}$S$_{0}$ level via coupling to a lossy cavity
mode.}%
\label{Fig3_Implementation}%
\end{figure}

\emph{The d-wave parent Hamiltonian} -- As a final remark, we note that the effective Hamiltonian above can be generalized to include a coherent interaction $V$,
\begin{eqnarray}
H_{\mathrm{eff}}=(V-\tfrac{\mathrm i}{2}\kappa)\sum_{i,\alpha}J^{\alpha\,\dag}_{i} J_{i}^\alpha.
\end{eqnarray}
For $\kappa\rightarrow0$ and interaction $V>0$ this
Hamiltonian can be identified as a \emph{parent Hamiltonian} \cite{auerbachbook} with
$|\mathrm{BCS}_{N}\rangle$  as unique stable ground state and gapped positive definite excitation spectrum. 
This parent Hamiltonian could be realised via a similar procedure to the induced dissipation, replacing the decay in step (iii) by induced interactions between atoms. This opens the possibility to use the d-wave state as an initial state for the preparation of the ground state of the Fermi-Hubbard model by a suitable adiabatic passage \cite{trebst}. Here, one can take advantage of the fact that (i) in the initial stages the system is protected by a gap $\sim 0.72V$, and (ii) the d-wave state has identical symmetry and similar energy to the conjectured Fermi-Hubbard ground state away from half filling. Thus, since no phase transition has to be crossed, a d-wave superfluid gap protection persists through the whole passage path.


We thank A. Gorshkov, K. Hammerer, B. Kraus and A. Kantian for discussions. This work was supported by the Austrian Science Fund through SFB F40 FOQUS and EUROQUAM\_DQS (I118-N16), and the EU through IP AQUTE.

\end{document}